# An overview of Intrusion Detection and Prevention Systems

Keturahlee Coulibaly
Faculty of Engineerimg and Informatics
Bradford University
Bradford, UK
Kcouliba@bradford.ac.uk

*Abstract*—Cyber threats are increasing not only in their volume but also in their sophistication and difficulty to detect. Attacks have become a national/global threat as they have targeted private and public, as well as government sectors over the years. This is a growing issue and organisations are taking steps to reduce, detect and prevent threats. To do this they need to use systems that are equipped with the capabilities to do either of those steps and develop them for the type of networks they use, for instance wired or wireless. One of these systems are Intrusion Detection Systems (IDS), which can be used as the first defense mechanism or a secondary defense mechanism of a threat or an attack. There are different types of attacks that can occur in a network, such as Denial of service (DoS)/Distributed Denial of Service (DDoS), port scanning, malware or ransomware and so forth that IDSs have a capability of detecting. Assisting in the mitigation of such attacks, there are also Intrusion Prevention Systems (IPS) whose role has a different purpose than that of IDSs. Unlike IDSs they not only detect threats but prevent them from disrupting the network, IPSs can be used in conjunction with IDSs to double the defenses. This paper provides an overview of IDS and their classifications and IPS. It will detail typical benefits and limitations to using IDSs, IPSs and the hybrids (such as Intrusions Detection Prevention Systems (IDPSs and more)) which will be discussed further. It will also outline developments in the making using ML and how it is used to improve these systems and the dilemmas they produce and possible ways to counter act them.

## I. INTRODUCTION

Different sectors of organisations have become victims of targeted cyber-attacks and thus far, attacks are on the rise and are being more advanced. Organisations are becoming more aware of the dangers and damages these security attacks can cause their businesses and infrastructure [1]. When they grow, they accumulate more data and sensitive information about others- such as clients or individuals and even government organisations. These can have devastating consequences and impacts on a company or organisations, this can be in the form of financial loss. Though "cyber-attacks are persistent" companies are identifying the attacks and or breaches less than before. On the other hand, those that can identify the threats and breaches are experiencing more attacks [2,9]. This infers that organisations who do not have the capabilities of identifying attacks as successfully as others may still be experiencing attacks or being part of a botnet without their knowledge [18,26]. Furthermore, with Advanced Persistent Threat (APT), the attack campaign can last for months or years exfiltrating the data without being detected [15,21,24]. Meaning that the attack frequency may be on par with organisations who are able to detect the high frequency of attacks. Statistics show that there has been a 60% increase in breaches or attacks in 'medium sized businesses, 61% in larger businesses and 52% high-income charities' than that of previous years, 32% of business and 22% of charities that would experience attacks [1]. However, what we know now about the failed detections these numbers could potentially be higher. This leads us to the topic of intrusion detection systems. What they are and how they can be used to help detect unwanted cyber-attacks. The topics remaining are as follows: Section two defines IDSs and the growing threats of cyber-attacks on Organisations. Section three will outline the various classifications of IDSs and their uses. Section four will examine the developments of Machine Learning (ML) in IDSs. Section five defines IPS and uses with IDSs.

## II. INTRUSION DETECTION SYSTEMS

Intrusion Detection Systems (IDSs) can assist organisations by preventing cyber-attacks and breaches from gaining access or further access into their network systems. An intrusion itself compromise and bypass the confidentiality, integrity and availability CIA of systems. With devices becoming widely accessible from the wireless connectivity, threats are likely to occur as wireless devices connected to the Internet of Things (IoTs) are easier to breach than wired network [3]. It is also said that the security and privacy of information are issues within IoTs, which further elaborates that the there is demand and need for IDSs, specifically in IoTs environments to mitigate attacks that happen to exploit vulnerabilities [10,4]. Organisations install security defenses to guarantee their network security, by installing systems such as IDSs, antivirus software, firewalls and so forth. However, these systems have limitations as well as drawbacks. For instance, firewalls protect against the unauthorised access, accessing private networks, they do not protect against viruses and or malware. Even IDSs have some drawbacks such the inability to process encrypted packets [13,5].

Intrusion detection systems do not have to be a stand-alone system. In fact, it can be used with other devices or even as a secondary defense mechanism with other devices like firewalls. For instance, if the firewalls dense is breached then the IDS can further defend the system by raising an alert to system administrators. Taking it a step further if and IDS was set up to be in conjunction with an Intrusion Prevention System (IPS) then the IPS can be alerted by the IDS and begin to mitigate the threat. Overall due to threats increasing, whether that's in the form of emails, or web attacks to ransomware "there is a need for strong, fast and reliable IDSs" [13].

IDSs typically, are used for detection of possible malicious threats. Acting as a "packet sniffer" by analyzing intercepted packets captured through "various communication mediums" (by different protocols between sender and receiver such as TCP/IP) [14]. There are many kinds of IDS that have been created which come with their many benefits. You can compare IDSs from their success rates in accurately detecting attacks, although IDSs have the capability to adapt as part of its feature if the developers wish to enhance the system if needs be. This leads to the next topic of discussion about IDSs.

### A. Intrussion Detection Systems Classifications

There are various kinds of IDSs which detect threats in different ways, there are Host-Based Intrusion Detection Systems (HIDS) and Network-Based Intrusion Detection Systems (NIDS). HIDSs are for single computers monitoring internal and external cyber-attacks [13]. Internal attacks refer to when systems identify when a security breach occurs, where programs access which resources by anyone or device within the organisation. External attacks on the other hand refer to when HIDS "analyses packets to and from a network", logging in activity and alerting administrators as well as alerting them to attacks that were successful or not. A passive system that waits for a possible threat to be an indicator for an attack about to occur before alerting administrators- it is not active in preventing the attack from happening. With HIDS security applications are installed on the system which monitors security, such as anti-virus and spyware, an example of an HIDS would be PortSentry [10]. However, due to the HIDS being installed on that single computer/host then they can only detect attacks that are not in another part of the network and have a high false positive rate [13].

NIDS can be strategically placed from the entry point and exit point of data coming and going through the network. In a network to detect malicious attacks capturing all the data going through the network. Raising an alert if a threat is indicated or when abnormal behaviour of network appears, an alert will also be sent [10].

### B. Intrusion Detection Techniques

In a NIDS there are two different types of techniques in indicating a threat, briefly mentioned previously, there are anomaly-based detections and signature based. Anomaly-based is where the information collected from traffic in the system comparing it to the gold standard for normal traffic/system behaviour [10]. When the system shows abnormal signs of behaviour it triggers an alert. Its advantage is that its capable of detecting intrusions that where unknown to the system previously-therefore, in a way detecting new attack patterns. However, due to the system been alerted to every irregularity, there are large amounts of false alarms in using this technique, it is also possible for some attacks to fall in normal behavioural standards to go undetected [14].

Unlike anomaly-based detections, signature-based detections base its alerts on known attack pattern signatures before alerting administrators, there are various algorithms used for the detection of attacks that can be used to yield better results, some of the popular NIDSs are SecureNet, Real Secure and Snort [10,11]. However, a drawback is that not all signatures of attacks may be established within this IDS, therefore, when attacks that do not fit within the pre-defined signatures occur, they will go undetected creating a vulnerability. Indicating that there is demand and need for IDSs to be able to detect new and unfamiliar attack signatures [16]. On the other hand, you can use both anomaly and signature-based techniques creating what is know as a hybrid; this increases the detection rate of the known signature attacks and decreases the false positive rate of unknown attacks. Creating hybrids allow for better accuracy rates overall and therefore, it is common for ML and/or Deep Learning models to use hybrids to improve [17,29].

## III. MACHINE LEARNING IN IDS

Companies are investing in studies to find ways in optimising the detection of attacks in networks in IDSs. As stated before, IDSs both anomaly and signature-based NIDSs have limitations and drawbacks in finding unknown signatures or anomalies of attacks or attacks that fall in in the baseline of normal behaviour that they go undetected. For instance, the IoTs have a higher risk of breaches from attacks, there has been development in using intelligent techniques and then comparing them with the one with the greater rate of accuracy, using machine learning [19].

### A. K-means Algorithm

There have been many developments of different types of algorithms to find which have better accuracy in detecting threats- such as k-means algorithm. It has been tested with the k-means algorithm with an anomaly-based IDS, by separating normal behavioural samples into clusters, then for the IDS to determine the difference between normal and abnormal "according to their distance from the clusters" of centroids using a "validated dataset a threshold value that is created". Indicating that any data received which are far from the cluster of centroids are identified as an abnormal [20]. Having a threshold where the normal sample stops and become abnormal can increase the accuracy of the IDS- in identifying some attacks that appear to be at the baseline of normal but in fact are not [12].

### B. Decision Trees

Another way ML can be used in NIDS is using a decision tree, using training and testing data sets to use- the training data is used to make the decision tree model and each leaf signifies any possibly outcomes. Classification models developed use the training data against the test data to classify malicious attacks in the data to measure the classifier against later/future data and not of past data, is an important aspect to its future accuracy. Observed it can be 99.9% accurate of true-positive account and false-positive at 0.1% using the latest datasets, showing great strides in development using ML. These are just a few of the various ways ML can be used to improve the accuracy of IDSs, particularly for NIDSs techniques such as anomaly-based IDS in reducing the high false positives [22].

IV. INTRUSION PREVENTION SYSTEMS (IPSS)

With the increased number of cyber-attacks IDSs are not the only systems in development to combat these threats. IPS have also been development, however their role is different to that of an IDS. While companies emphasise the use of IDSs, they are unable to prevent all threats, therefore it is known for companies to use IDSs alongside firewalls and/or antivirus software as previously discussed. On the other hand, as also previously discussed, these other systems have their own limitations and drawbacks- ultimately they must be optimal configuration with IDSs in order to operate at joint functionality eliminating any negative effects of joining implementations, such as conflicts and time delays with linkage [23].

Whereas IPSs provide popular detection and prevention flexibility in controls, which for instance create the opportunity to resolve time delays which IDSs can incur with other security systems. Implying it "having the capabilities of IDSs" as it can also not only detect intrusions but also take counter measures to prevent/mitigate them, this can be for instance with attack pattern recognition [23]. Therefore, this system can also be called Intrusions Detection Prevention Systems (IDPSs). When working with wireless networks a wireless-based IDPS can analyse the traffic and "detects any unauthorised wireless local area network use and takes the necessary steps to counter them". However, it is not capable of detecting suspicious behaviour in application and transport layers or protocol activities, it is placed in a range where it can be monitored in the wireless network [25]. In the same way NIDPSs like NIDSs monitored network traffic, searching for malicious activity but could not prevent unfamiliar attacks. On the other hand, host-based IDPSs monitors a single hosts system and "can prevent system level attacks and detect attacks that the NIDPS are incapable of detecting- such as identifying unusual traffic, like Distributed Denial of Service attacks (DDoS), malware (e.g. Worms), even policy violations [25].

The overall purpose of IDPSs is to monitor systems and protect the network against intruders, then provide a report to administrators if there are attacks that occurred in the networks environment. While IDS typically protect against outsider attacks, it makes it harder for them to detect insider attacks, this is due to Intrusions Preventions Systems (IPS) being placed at the edge of networks and the concerns are that attacks can still occur from the inside before reaching the IPS. Therefore, further development to revolve that dilemma still need improvement-potentially by integrating other systems with IPS that can communicate to achieve solutions to such dilemmas. However, IPSs have more capabilities overall than IDSs alone, they are considered to be an enhancement [27,28].

*A. IPS effectivness*

For IPSs to detect and prevent, there are various measures to compare and measure the effectiveness of IPSs and because IPS are important aspects of security, the choices it makes are crucial and can be on the bases of the IPSs characteristics which would be "the distributive property, autonomy communication, cooperation, responsiveness and adaptability, measuring the effectiveness" by [28]:

1) Testing the rate of false positives and false negatives.
2) response time in an overloaded network environment.
3) The possible capabilities of updating the database of signatures or even to modify the signatures.
4) Having a need for majority of audit data.

Since IPSs mainly consist of a single block that will handle an entire analysis on the network, it still has constraints, such as, to name a few having difficulty updating and a consumption of system resources. In overcoming these drawbacks new design strategies are taking place to revolve them. One of the current strategies are to have a model composed of four blocks instead of one- "source of information, sensor, analyser and manager this is the development of the Defense Advanced Research Projects Agency (DARPA) committee and is generally used as a standard in a majority of new current IPS [28].

The study of IDSs has provided a realization of the importance of the position played by IPS, the Host Based IPS and the Network-Based IPS in security. IPSs must meet specific requirements, the characteristics "should be primarily chosen typically by the need and the security hardware and software constraints". The types of IPSs in use can be determined by the "location of the IPS, frequency of use, the method of detection and the response of the IPS". A hybrid of IPSs using HIPSs and NIPSs can also be used if considering those factors and characteristics as they can be distributed on several machines analyzing data from the different sources and assessing the "information from two sources and immune system". Therefore, a 'clonal theory' could potentially generate more detectors from an attack pattern and recognise not only the attack but its variants too, including other similar attacks, this is one of the ideal scenarios which is still in development [28,30].

V. CONCLUSION

Overall in this overview we have seen the benefits of IDSs and IPSs and their general benefits to organisations in preventing cyber-attacks in their systems. However, there has been some insight into their limitations and drawbacks as a standalone system, which can cause organisations to source further resources, however, this may provide room for developments. On the other hand, some of the studies demonstrate the use of hybrid systems like the IDPSs used as a counter measure for the drawbacks of their standalone systems, like IDS and IPS, which are promising and they counter acted some of the flaws each system had when used on their own. As the development of ML are used to incorporate into these systems hybrid or not, they have increased accuracy of detections of attacks in systems, with different algorithms for different in-house systems which yielded promising results in those studies. In conclusion using hybrid systems and integrating ML into them has reduced the amount of false alarms in IDSs, IPSs and IDPSs overall. Whether there will be an optimal system with 0% false alarms and 100% detection of threats, whether abnormal behavior of attacks or different variants of attack signatures, are still in development and potentially ML will be one of the key factors in making that a reality if the idea is at all plausible. This this because each adaptation of an IDS, IPS or IDPS there has still been limitations in the system that cause for vulnerability therefore, there are still various theories in development into improving those dilemmas.

REFERENCES

[1] Department for digital, Culture, Media and Sport (2019) "Cyber Security Breaches Surcey Statistical release"

[2] I. Ghafir and V. Prenosil. "Proposed Approach for Targeted Attacks Detection," Advanced Computer and Communication Engineering Technology, Lecture Notes in Electrical Engineering. Phuket: Springer International Publishing, vol. 362, pp. 73-80, 9, 2016.

[3] Liao HJ, Lin CH, Lin YC, Tung KY. Intrusion detection system: A comprehensive review. Journal of Network and Computer Applications. 2013 Jan 1;36(1):16-24.

[4] U. Raza, J. Lomax, I. Ghafir, R. Kharel and B. Whiteside, "An IoT and Business Processes Based Approach for the Monitoring and Control of High Value-Added Manufacturing Processes," International Conference on Future Networks and Distributed Systems. Cambridge, United Kingdom, 2017.

[5] I. Ghafir, M. Husak and V. Prenosil, "A Survey on Intrusion Detection and Prevention Systems," IEEE/UREL conference, Zvule, Czech Republic, pp. 10-14, 2014.

[6] Elrawy MF, Awad AI, Hamed HF. Intrusion detection systems for IoT-based smart environments: a survey. Journal of Cloud Computing. 2018 Dec 1;7(1):21.

[7] AlYousef MY, Abdelmajeed NT. Dynamically Detecting Security Threats and Updating a Signature-Based Intrusion Detection System's Database. Procedia Computer Science. 2019 Jan 1;159:1507-16.

[8] Innella P. The evolution of intrusion detection systems. SecurityFocus, November. 2001 Nov 16;16.

[9] I. Ghafir and V. Prenosil, "Advanced Persistent Threat and Spear Phishing Emails." International Conference Distance Learning, Simulation and Communication. Brno, Czech Republic, pp. 34-41, 2015.

[10] Pharate A, Bhat H, Shilimkar V, Mhetre N. Classification of Intrusion Detection System. International Journal of Computer Applications. 2015 Jan 1;118(7).

[11] J. Svoboda, I. Ghafir, V. Prenosil, "Network Monitoring Approaches: An Overview," International Journal of Advances in Computer Networks and Its Security (IJCNS), vol. 5(2), pp. 88-93, 2015.

[12] I. Ghafir and V. Prenosil, "DNS query failure and algorithmically generated domain-flux detection," International Conference on Frontiers of Communications, Networks and Applications. Kuala Lumpur, Malaysia, pp. 1-5, 2014.

[13] Mehmood T, Rais HB. Machine learning algorithms in context of intrusion detection. In2016 3rd International Conference on Computer and Information Sciences (ICCOINS) 2016 Aug 15 (pp. 369-373). IEEE.

[14] Jyothsna V, Prasad KM. Anomaly-Based Intrusion Detection System. InComputer and Network Security 2019 Jun 11. IntechOpen.

[15] I. Ghafir, V. Prenosil, M. Hammoudeh and U. Raza, "Malicious SSL Certificate Detection: A Step Towards Advanced Persistent Threat Defence," International Conference on Future Networks and Distributed Systems. Cambridge, United Kingdom, 2017.

[16] Bronte R, Shahriar H, Haddad HM. A signature-based intrusion detection system for web applications based on genetic algorithm. InProceedings of the 9th International Conference on Security of Information and Networks 2016 Jul 20 (pp. 32-39). ACM.

[17] Jamadar R, Ingale S, Panhalkar A, Kakade A, Shinde M. Survey of Deep Learning Based Intrusion Detection Systems for Cyber Security.

[18] I. Ghafir, V. Prenosil, and M. Hammoudeh, "Botnet Command and Control Traffic Detection Challenges: A Correlation-based Solution." International Journal of Advances in Computer Networks and Its Security (IJCNS), vol. 7(2), pp. 27-31, 2017.

[19] da Costa KA, Papa JP, Lisboa CO, Munoz R, de Albuquerque VH. Internet of Things: A survey on machine learning-based intrusion detection approaches. Computer Networks. 2019 Mar 14;151:147-57.

[20] KarsligEl ME, Yavuz AG, Güvensan MA, Hanifi K, Bank H. Network intrusion detection using machine learning anomaly detection algorithms. In2017 25th Signal Processing and Communications Applications Conference (SIU) 2017 May 15 (pp. 1-4). IEEE.

[21] I. Ghafir and V. Prenosil, "Malicious File Hash Detection and Drive-by Download Attacks," International Conference on Computer and Communication Technologies, series Advances in Intelligent Systems and Computing. Hyderabad: Springer, vol. 379, pp. 661-669, 2016.

[22] Jamadar, Riyazahmes A. Network intrusion detection system using machine learning. 2018 Dec 01 11(48): 1-6

[23] Cai C, Mei S, Zhong W. Configuration of intrusion prevention systems based on a legal user: the case for using intrusion prevention systems instead of intrusion detection systems. Information Technology and Management. 2019:1-7.

[24] I. Ghafir, J. Svoboda and V. Prenosil, "Tor-based malware and Tor connection detection," International Conference on Frontiers of Communications, Networks and Applications. Kuala Lumpur, Malaysia, pp. 1-6, 2014.

[25] Tasneem A, Kumar A, Sharma S. Intrusion Detection Prevention System using SNORT. International Journal of Computer Applications.;975:8887.

[26] I. Ghafir, J. Svoboda, V. Prenosil, "A Survey on Botnet Command and Control Traffic Detection," International Journal of Advances in Computer Networks and Its Security (IJCNS), vol. 5(2), pp. 75-80, 2015.

[27] Putra AS, Surantha N. Internal Threat Defense using Network Access Control and Intrusion Prevention System.

[28] Farhaoui Y. How to secure web servers by the intrusion prevention system (IPS)?. International Journal of Advanced Computer Research. 2016 Mar 1;6(23):65.

[29] I. Ghafir, V. Prenosil, M. Hammoudeh, F. J. Aparicio-Navarro, K. Rabie and A. Jabban, "Disguised Executable Files in Spear-Phishing Emails: Detecting the Point of Entry in Advanced Persistent Threat." International Conference on Future Networks and Distributed Systems. Amman, Jordan, 2018.

[30] Parthasarathy K. Clonal selection method for immunity-based intrusion detection system. Project Report. 2014:1-9.